\documentclass[twocolumn,preprintnumbers,a4paper,
superscriptaddress,floatfix,twoside,prd]{revtex4}

\usepackage{multirow}
\usepackage{bm}
\usepackage{epsfig}
\usepackage{graphics}
\usepackage{natbib}

\usepackage{fancyh}
\usepackage[l]{floatflt}
\usepackage{epsfig}
\usepackage{amssymb}
\usepackage{latexsym}
\usepackage{times}
\usepackage{slashed}
\usepackage{upgreek}
\usepackage{amsmath}
\usepackage{amsfonts}
\usepackage{amsbsy}
\usepackage{amscd}
\usepackage{bbm}

\newcommand{\fref}[1]{Fig.~\ref{#1}}

\newcommand{\Eref}[1]{Eq.~(\ref{#1})}

\newcommand{\Sref}[1]{Sec.~\ref{#1}}
\newcommand{\Fref}[1]{Fig.~\ref{#1}}
\newcommand{\Tref}[1]{Table~\ref{#1}}
\newcommand{\cref}[1]{Ref.~\cite{#1}}

\newcommand{\hepph}[1]{{\ftn\tt hep-ph/#1}}

\newcommand{\arxiv}[1]{{\ftn\tt  arXiv:#1}}

\newcommand{\bal}{\begin{align}}
\newcommand{\eal}{\end{align}}
\newcommand{\beqs}{\begin{subequations}}
\newcommand{\eeqs}{\end{subequations}}
\newcommand{\eec}{\end{center}}
\newcommand{\bec}{\begin{center}}

\newcommand{\ecs}{\end{cases}}
\newcommand{\bcs}{\begin{cases}}

\newcommand{\eem}{\end{matrix}}
\newcommand{\bem}{\begin{matrix}}
\newcommand{\eeq}{\end{equation}}
\newcommand{\beq}{\begin{equation}}
\newcommand{\ba}{\begin{array}}
\newcommand{\ea}{\end{array}}
\newcommand{\bea}{\begin{eqnarray}}
\newcommand{\eea}{\end{eqnarray}}
\newcommand{\baq}{\begin{eqnarray}}
\newcommand{\eaq}{\end{eqnarray}}

\newcommand\eqs[2]{Eqs.~(\ref{#1}) and (\ref{#2})}

\newcommand\eqss[3]{Eqs.~(\ref{#1}), (\ref{#2}) and (\ref{#3})}
\newcommand\eqsss[4]{Eqs.~(\ref{#1}), (\ref{#2}), (\ref{#3}) and (\ref{#4})}
\newcommand\eqssss[5]{Eqs.~(\ref{#1}), (\ref{#2}), (\ref{#3}), (\ref{#4}) and (\ref{#5})}

\newcommand{\ftn}{\footnotesize}

\newcommand{\TeV}{{\mbox{\rm TeV}}}

\newcommand{\GeV}{{\mbox{\rm GeV}}}

\newcommand{\ZeV}{{\mbox{\rm ZeV}}}
\newcommand{\EeV}{{\mbox{\rm EeV}}}
\newcommand{\PeV}{{\mbox{\rm PeV}}}

\newcommand{\etal}{{\it et al.\/}}

\def\to{\rightarrow}

\def\llgm{\left\lgroup}
\def\rrgm{\right\rgroup}
\def\lf{\left(}
\def\rg{\right)}
\newcommand\vev[1]{\langle {#1} \rangle}
\newcommand\vevi[1]{\langle {#1} \rangle_{\rm I}}
\newcommand{\Gr}{\ensuremath{\widetilde{G}}}
\newcommand{\Yb}{\ensuremath{Y_{B}}}
\newcommand{\Yg}{\ensuremath{Y_{3/2}}}
\newcommand{\Vhi}{\ensuremath{ V_{\rm HI}}}

\newcommand{\Hhi}{\ensuremath{ H_{\rm HI}}}
\newcommand{\what}{\ensuremath{\widehat}}
\newcommand{\wtilde}{\ensuremath{\widetilde}}
\newcommand{\Khi}{\ensuremath{K}}

\newcommand{\Vhio}{\ensuremath{ V_{\rm I0}}}
\newcommand{\Ns}{\ensuremath{{N_\star}}}

\newcommand{\mP}{\ensuremath{m_{\rm P}}}
\newcommand{\Mgut}{\ensuremath{M_{\rm GUT}}}

\newcommand{\Ggut}{\ensuremath{\mathbb{G}}}

\newcommand{\Gbl}{\ensuremath{\mathbb{G}_{B-L}}}
\newcommand{\Gsm}{\ensuremath{\mathbb{G}_{\rm SM}}}

\newcommand{\lm}{\ensuremath{\lambda_\mu}}

\def\openone{\leavevmode\hbox{\small1\kern-3.8pt\normalsize1}}
\newcommand{\dV}{\ensuremath{\Delta V_{\rm HI}}}

\newcommand{\fp}{\ensuremath{f_{p}}}

\newcommand{\ft}{\ensuremath{f_{\rm T}}}

\newcommand{\kb}{\ensuremath{K_2}}
\newcommand{\ka}{\ensuremath{K_1}}

\newcommand{\tki}{\ensuremath{\widetilde K_\ell}}
\newcommand{\kst}{\ensuremath{K_{\rm st}}}
\newcommand{\tkbs}{\ensuremath{\widetilde K_{\rm 2st}}}
\newcommand{\tkas}{\ensuremath{\widetilde K_{\rm 1st}}}
\newcommand{\tkis}{\ensuremath{\widetilde K_{\ell\rm st}}}

\newcommand{\nst}{\ensuremath{N_{\rm st}}}
\newcommand{\nb}{\ensuremath{N}}
\newcommand{\na}{\ensuremath{N}}
\newcommand{\nmin}{\ensuremath{N_{\rm min}}}

\newcommand{\Gsn}{\ensuremath{\what{\Gamma}_{\rm \dph}}}
\newcommand{\GNsn}{\ensuremath{\what{\Gamma}_{\dph\to N_i^c}}}
\newcommand{\Ghsn}{\ensuremath{\what{\Gamma}_{\dph\to H}}}

\newcommand{\msn}{\ensuremath{\what m_{\rm \dph}}}
\newcommand{\aS}{\ensuremath{{\rm a}_S}}
\newcommand{\Ald}{\ensuremath{A_\lambda}}

\newcommand{\hd}{{\ensuremath{H_d}}}
\newcommand{\hu}{{\ensuremath{H_u}}}
\newcommand{\ssni}{\ensuremath{\widetilde N^c_i}}
\newcommand{\sni}{\ensuremath{N^c_i}}

\newcommand{\ks}{\ensuremath{k_\star}}
\newcommand{\ns}{\ensuremath{n_{\rm s}}}
\newcommand{\as}{\ensuremath{a_{\rm s}}}
\newcommand{\As}{\ensuremath{A_{\rm s}}}

\newcommand{\rcc}{\ensuremath{\mathcal{R}}}

\newcommand{\Ve}{\ensuremath{ V}}

\newcommand{\dphi}{\ensuremath{\what{\delta\phi}}}
\newcommand{\dph}{\ensuremath{\delta\phi}}

\def\ve{\varepsilon}

\def\bbet{{\bar\beta}}
\def\al{{\alpha}}

\def\bt{{\beta}}

\def\th{{\theta}}
\def\thb{{\bar\theta}}

\def\thn{{\theta_{\Phi}}}

\newcommand{\Trh}{\ensuremath{T_{\rm rh}}}
\newcommand{\sg}{\ensuremath{\phi}}

\newcommand{\sgx}{\ensuremath{\phi_\star}}
\newcommand{\sgf}{\ensuremath{\phi_{\rm f}}}

\newcommand{\ld}{\ensuremath{\lambda}}
\newcommand{\ldu}{\ensuremath{\uplambda}}
\newcommand{\Ld}{\ensuremath{\Lambda}}
\newcommand{\kp}{\ensuremath{\kappa}}

\newcommand{\se}{\ensuremath{\widehat \phi}}

\newcommand{\geu}{\ensuremath{ g}}

\newcommand{\mgr}{\ensuremath{m_{3/2}}}
\newcommand{\mmgr}{\ensuremath{\mu/m_{3/2}}}

\newcommand{\dK}{\ensuremath{\Delta K}}
\newcommand{\dW}{\ensuremath{\Delta W}}

\newcommand{\Dex}{\ensuremath{\Delta_{\star}}}

\newcommand{\vtau}{\ensuremath{\uptau}}

\newcommand{\phc}{\ensuremath{\Phi}}
\newcommand{\phcb}{\ensuremath{\bar\Phi}}
\newcommand{\wrh}{\ensuremath{w_{\rm rh}}}

\newcommand\mtta[4]{\mbox{
$\llgm\bem #1 &#2 \cr #3& #4\eem\rrgm$}}

\newcommand{\am}{\ensuremath{{\rm a}_{\mu}}}

\newcommand{\rhna}{\ensuremath{N^c_1}}
\newcommand{\rhni}{\ensuremath{N^c_i}}
\newcommand{\mrh[1]}{\ensuremath{M_{#1N^c}}}
\newcommand{\lrh[1]}{\ensuremath{\lambda_{#1N^c}}}

\newcommand{\mntau}{\ensuremath{m_{\nu_\tau}}}

\newcommand{\mrha}{\ensuremath{M_{1N^c}}}

\def\Ka{K\"{a}hler potential}
\def\Km{K\"{a}hler manifold}
\def\Kaa{K\"{a}hler~}

\newcommand{\plk}{{\it Planck}}

\def\tpmd{{T$_p$-Model}}

\def\fhi{{$p$HI}}

\def\actc{{\sf\small P-ACT-LB-BK18}}
\def\actcf{{\sf\ftn P-ACT-LB-BK18}}

\newcommand{\diag}{\ensuremath{{\sf diag}}}

\renewcommand{\refname}{{\bf\scshape References}}

\renewcommand{\thesubsection}{{\small\sf\Alph{subsection}}}

\newcommand{\sgm}{\ensuremath{\phi_{\rm mx}}}
\newcommand{\Vm}{\ensuremath{V_{\rm mx}}}

\newcommand{\rhofi}{{\ensuremath{\rho_{{\rm i}}}}}

\newcommand{\rhoR}{{\ensuremath{\rho_{\rm R}}}}
\newcommand{\rhoRi}{{\ensuremath{\rho_{\rm Ri}}}}

\newcommand{\Ti}{{\ensuremath{T_{\rm i}}}}

\newcommand{\vtrh}{\ensuremath{\vtau_{\rm rh}}}

\renewenvironment{subequations}{%
\refstepcounter{equation}%
\setcounter{parentequation}{\value{equation}}%
  \setcounter{equation}{0}
  \def\theequation{\theparentequation{\sf\small\alph{equation}}}%
  \ignorespaces
}{%
  \setcounter{equation}{\value{parentequation}}%
  \ignorespacesafterend
}

\begin{document}



\title{\bf\scshape Updating GUT-Scale Pole Higgs Inflation After ACT DR6}

\author{\scshape Constantinos Pallis\\ {\it School of Technology,  Aristotle University of
Thessaloniki, Thessaloniki, GR-541 24 GREECE} \\  {\sl e-mail
address: }{\ftn\tt kpallis@auth.gr}}



\begin{abstract}

\noindent {\ftn \bf\scshape Abstract:} We consider models of
chaotic inflation driven by the real parts of a conjugate pair of
Higgs superfields involved in the spontaneous breaking of a grand
unification symmetry at a scale assuming its value within MSSM. We
combine a superpotential, which is uniquely determined by applying
a continuous $R$ symmetry, with two fractional shift-symmetric \Ka
s introducing two free parameters $(p,N)$. The inflationary
observables provide an excellent match to the recent ACT data for
$1.355\leq p\leq6.7$ and $6\cdot10^{-5}\leq N\leq0.7$. The
attainment of inflation allows for subplanckian inflaton values
and possibly detectable primordial gravitational waves with
$(p,N)$ values of order unity. A solution to the $\mu$ problem of
MSSM and baryogenesis via non-thermal leptogenesis can be also
accommodated by embedding the model into a $B-L$ extension of
MSSM.
\\ \\ {\scriptsize {\sf PACs numbers: 98.80.Cq, 04.50.Kd, 12.60.Jv, 04.65.+e}
\hfill {\sl\bfseries Published in} {\sl Phys. Rev. D} {\bf 113},
no.~1, 015033 (2026)}

\end{abstract}\pagestyle{fancyplain}

\maketitle

\rhead[\fancyplain{}{ \bf \thepage}]{\fancyplain{}{\sl Updating
GUT-Scale Pole HI After ACT DR6}} \lhead[\fancyplain{}{\sl C.
Pallis}]{\fancyplain{}{\bf \thepage}} \cfoot{}

\section{Introduction} \label{intro}

The announcement of \emph{Data Release 6} ({\sf\ftn DR6}) from the
\emph{Atacama Cosmology Telescope} ({\sf\ftn ACT})
\cite{act,actin} fuelled a plethora of works
\cite{act0,act2,act4,gup,oxf,indi,kina,actlee,rhc,rhb,
rha,act5,nmact,maity,reh8, actattr,act1,actj,yin,actpal,
act3,act6,act7,act8,actellis,acttamv,ketov,
r2a,r2b,r2drees,r2mans,r2li,heavy1,heavy,fhi1,fhi2,fhi3,fhi4,fhi5,smth}
which revise many well-motivated inflationary models -- for a
review see \cref{review}. This is because the combination of the
aforementioned data with the \emph{cosmic microwave background}
({\sf\small CMB} measurements by \plk\ \cite{plin} and
\emph{BICEP/Keck} ({\sf\small BK}) \cite{bcp}, together with the
\emph{Dark Energy Spectroscopic Instrument} ({\sf\small DESI})
{\it Baryon Acoustic Oscillation} ({\sf\small BAO}) results
\cite{desi} suggests a value of the (scalar) spectral index \ns\
substantially larger than the one indicated from the \plk\ data
\cite{plin}. Namely, taking into account also the tensor-to-scalar
ratio $r$, the so-called \actc\ data entails \cite{actin}
\beq \label{data} \ns=0.9743\pm0.0068\>\>\mbox{and}\>\>r\leq0.038
\eeq
at 95$\%$ \emph{confidence level} ({\sf\small c.l.}).

In one recent paper \cite{actpole} we propose a variant of pole
inflation \cite{terada,pole,pole1,sor,tmhi,prl} which assures
compatibility with the $(\ns,r)$ values in \Eref{data}. This goal
becomes possible by introducing a particular class of rational \Ka
s $K$ which include one extra parameter $p$ compared to the
$\alpha$- ($N$- in our notation) attractor models
\cite{alinde,eno7,tmodel}. Most notably, this achievement is
independent of the type of the monomial superpotential $W$
selected leading, thereby, to the establishment of a new set of
$(N,p)$ attractors. Working in the context of \emph{Supergravity}
({\sf\ftn SUGRA}), in our present investigation  we adapt our
proposal in \cref{actpole} to a \emph{Grand Unified Theory} ({\sf
\ftn GUT}) including the inflaton in a GUT-scale Higgs sector. In
a such case, both $W$ and $K$ are consistent with a gauge symmetry
rendering our scheme much more well-defined. Following the
terminology of \cref{actpole} we here realize \tpmd\ inflation
identifying the inflaton with the real part of a conjugate pair of
Higgs superfields -- cf. \cref{sor, tmhi}. We therefore could name
our model \tpmd\ \emph{Higgs Inflation} ({\sf\small HI}) or, for
short, {\sf\small \fhi} -- for other works on GUT-scale HI see
e.g. \cref{hi1, hi2, hi3, hi4, hi5, nmBL, jhep,lrcs,ighi}.

Apart from the compatibility with \Eref{data} \fhi\ offers high
enough $r$ values which can be hopefully tested in the near future
\cite{det}. Nonetheless, it also allows us to explore consequences
beyond the inflationary phenomenology. Namely, embedding \fhi\
into a $B-L$ extension of \emph{Minimal Supersymmetric Standard
Model} ({\sf\small MSSM}) we can accommodate a resolution of the
$\mu$ problem, following the mechanism in \cref{dvali, tmhi}, and
acceptable baryogenesis via \emph{non-thermal leptogenesis}
({\sf\small nTL}) \cite{dreeslept,zhang,lept} consistently with
the gravitino ($\Gr$) constraint \cite{brand,kohri}. The success
of our post-inflationary scenario bounds from above the ratio
$\mmgr$ -- where $\mgr$ is the $\Gr$ mass  -- and so it gives us
the opportunity to link the high- to low-scale
\emph{supersymmetry} ({\sf \small SUSY}).


Below, in \Sref{set}, we describe the building blocks ($W$ and
$K$'s) of our proposal, we outline the derivation of the
inflationary potential in \Sref{infv} and test our model against
observations in \Sref{res}. Finally, we expose a post-inflationary
completion in \Sref{post} and summarize our conclusions in
\Sref{con}. In Appendix A we present our approach to the reheating
stage in our model. Unless otherwise stated, we use units where
the reduced Planck scale $\mP = 2.43\cdot 10^{18}~\GeV$ is equal
to unity.


\section{Model Set-up} \label{set}

Our starting point is the superpotential known from the models of
F-term hybrid inflation \cite{fhi1,fhi2,fhi3,fhi4,fhi5}
\beq W=\ld S\lf\bar\Phi\Phi-M^2/2\rg\label{whi} \eeq
which is uniquely determined, at renormalizable level, by a gauge
symmetry $\Ggut$ and a convenient continuous $R$ symmetry. Here,
$\ld$ and $M$ are two constants which can both be taken positive
by field redefinitions; $S$ is a left-handed superfield, singlet
under $\Ggut$; $\bar\Phi$ and $\Phi$ is a pair of left-handed
superfields belonging to non-trivial conjugate representations of
$\Ggut$, and reducing its rank by their \emph{vacuum expectation
values} ({\sf\small v.e.vs}). Just for definiteness we restrict
ourselves to $\Gbl=\Gsm\times U(1)_{B-L}$ which consists the
simplest GUT beyond the MSSM  -- here  $\Gsm$ is the gauge group
of MSSM and $B$ and $L$ denote the baryon and lepton number. The
$B-L$ and $R$ charges of $S, \phcb$ and $\Phi$ can be assigned as
shown in Table 1 of \cref{nmBL}. At the SUSY limit  $W$ leads to a
$B-L$ phase transition since, at the SUSY vacuum
\beq
\label{vevs}\vev{S}\simeq0~~\mbox{and}~~\vev{\Phi}=\vev{\bar\Phi}=
M/\sqrt{2}~~\mbox{for}~~M\ll1,\eeq
%
$U(1)_{B-L}$ is spontaneously broken via $\vev{\Phi}$ and
$\vev{\bar\Phi}$.

To realize \fhi\ using $W$ in \Eref{whi} -- with  $S$ stabilized
at the origin and the inflaton included in the $\phcb-\phc$ system
-- a careful choice of $K$ is imperative as in
\cref{sor,jhep,nmBL}. The proposed here $K$ includes two
contributions without mixing between $\phcb-\Phi$ and $S$, i.e.,
\beq \wtilde K_{\ell\rm st}= \wtilde
K_\ell+\kst~~\mbox{with}~~\ell=1,2 \label{tkis}\eeq
from which $\kst$ successfully stabilizes $S$ along the
inflationary path -- see below -- without invoking higher order
terms. We adopt the form \cite{su11}
\beq \kst=\nst\ln\lf1+{|S|^2/\nst}\rg~~\mbox{with}~~0<\nst<6
\label{kst}\eeq
which parameterizes the compact manifold $SU(2)/U(1)$ with
curvature $2/\nst$. On the other hand,  $\wtilde K_\ell$ is
devoted to the inflationary part and includes two contributions: a
real one $K_\ell$ which determine the \Kaa\ metric for the
inflaton sector and an holomorphic (and anti-holomorphic) part
$K_{\rm sh}$ which provide $\wtilde K_\ell$ with a shift symmetry
assuring $\tki=0$ during \fhi\ -- cf. \cref{actpole}. Therefore,
$\wtilde K_\ell$ has the structure
\beq \wtilde K_\ell=K_\ell+K_{\rm
sh}~~\mbox{with}~~\ell=1,2.\label{tki}\eeq

The key ingredient of our proposal is the selection of a totally
fractional $K_\ell$ for $\ell=1$, which reads
\beqs\beq \ka=\frac{\nb}{\left(1-|\phc|^2-|\phcb|^{2}\right)^p}
\label{ka} \eeq
defined for $\na>0$ and $|\phc|^2+|\phcb|^{2}<1$. Identical
results can be achieved if we select for $\ell=2$
\beq \kb=\frac{\na}{2\left(1-2|\phc|^2\right)^p}+\frac{\na}{2\left(1-2|\phcb|^2\right)^p}\\
\label{kb}\eeq\eeqs
defined for $\na>0$, $|\phc|<1/\sqrt{2}$ and $|\phcb|<1/\sqrt{2}$.
Although $\kb$ seems to be more contrived than $\ka$ -- due to the
appearance of the same factor in the numerators and exponent in
the denominators -- both $K$'s share the same symmetries: These
are invariant under $\Ggut$ and the interchange
$\phc\leftrightarrow\phcb$. As regards the parameters $(p,N)$ we
expect that $N$ is constrained observationally -- as for the $N$
attractors \cite{alinde, tmhi, sor} -- and confine ourselves
conservatively in the range $0.1\leq p\leq10$.

On the other hand, $K_{\rm sh}$ is common for both $\wtilde
K_\ell$ and takes the form
\beq K_{\rm
sh}=-\frac{\na}{2\left(1-2\phcb\phc\right)^p}-\frac{\na}{2\left(1-2\phcb^*\phc^{*}\right)^p},\label{ksh}\eeq
where star denotes complex conjugation. It is obvious that $K_{\rm
sh}$ let unchanged the \Kaa metric by construction. Actually, the
function in the denominators of $K_{\rm sh}$ is the unique
possible quadratic function consistent with the gauge symmetry --
cf. \Eref{whi}. Both $K$'s in \Eref{tki} parameterize hyperbolic
\Km s but without constant curvatures unlike the case of T-Model
HI \cite{sor}.

\section{Inflationary Potential}\label{infv}

The appropriateness of the proposed $W$ and $\tkis$ in
\eqs{whi}{tkis} can be verified if we identify the inflationary
potential and the canonically normalized inflaton with those
introduced in \cref{actpole} for \tpmd\ inflation.

To embark on it, we focus on the SUGRA lagrangian density for the
complex scalar fields $z^\al=S,\phc,\phcb$ -- denoted by the same
superfield symbol -- which can be written as \cite{linde1}
\beq\label{action1} {\cal L} = \sqrt{-
\mathfrak{g}}\lf-\frac{1}{2}\rcc +K_{\al\bbet} \geu^{\mu\nu}D_\mu
z^\al D_\nu z^{*\bbet}-\Ve\rg, \eeq
where $\mathfrak{g}$ is the determinant of the background
Friedmann-Robertson-Walker metric $g^{\mu\nu}$ with signature
$(+,-,-,-)$, $\rcc$ is the Ricci scalar. The \Kaa metric
$K_{\al\bbet}$ and its inverse $K^{\bbet\al}$ are defined as
follows
\beq \label{kdef} K_{\al\bbet}={K_{,z^\al
z^{*\bbet}}}>0\>\>\>\mbox{with}\>\>\>K^{\bbet\al}K_{\al\bar
\gamma}=\delta^\bbet_{\bar\gamma},\eeq
where $z^\al$ are complex scalar fields -- the symbol $,z^\al$ as
subscript denotes derivation \emph{with respect to} ({\ftn\sf
w.r.t}) $z^\al$ -- and $D_\mu$ is the gauge covariant derivative.
Also $\Ve$ is the SUGRA potential which can be found in terms of
$W$ and $\tkis$ in \eqs{whi}{tkis} via the formula
\beqs\beq \Ve=e^{\Khi}\left(K^{\al\bbet}{\rm F}_\al {\rm
F}^*_\bbet -3{\vert W\vert^2}\right)+\frac{g^2}2 \mbox{$\sum_a$}
{\rm D}_a {\rm D}_a,\label{Vsugra} \eeq
where the F and D terms are defined as follows
\beq {\rm F}_\al =W_{,z^\al} +K_{,z^\al}W ~~\mbox{and}~~{\rm
D}_a=z^\al\lf T_a\rg^\bt_\al K_\bt \eeq\eeqs
with  $K_{\al}={\Khi_{,z^\al}}$ and the summation is applied over
the generators $T_a$ of $\Ggut$. If we parameterize $\Phi,
\bar\Phi$ and $S$ as follows
\beq\label{hpar} \Phi={\sg e^{i\th}} \cos\thn,~~\bar\Phi={\sg
e^{i\thb}} \sin\thn~~\mbox{and}~~S=({s +i\bar
s})/{\sqrt{2}}\,,\eeq
with $0\leq\thn\leq\pi/2$, we can easily deduce from \Eref{Vsugra}
that a D-flat direction occurs at
\beq \label{inftr} \vevi{\bar
s}=\vevi{s}=\vevi{\th}=\vevi{\thb}=0\>\>\>\mbox{and}\>\>\>\vevi{\thn}={\pi/4},\eeq
where the symbol $\vevi{Q}$ means the value of the quantity $Q$
during \fhi. Since $\vevi{K}=0$ and $\vevi{K_{SS^*}}=1$ for both
$\tkis$ in \Eref{tkis} the only surviving term from $V$ in
\Eref{Vsugra} is
\beq \label{Vhio}\Vhi=\vevi{V}=\vevi{e^{K}K^{SS^*}\,
|W_{,S}|^2}\,={\ld^2(\sg^2-M^2)^2}/{4}\eeq
which reduces to the quartic potential considered in
\cref{actpole} for $M\ll1$ -- see below.

To specify the canonically normalized inflaton, we note that, for
both $\tkis$ in \Eref{tkis}, $K_{\al\bbet}$ along the
configuration in \Eref{inftr} takes the form
\beq \label{kab} \vevi{K_{\al\bbet}}=\diag\lf
M_{\phcb\phc},\vevi{K_{SS^*}}\rg\eeq
where the matrix $M_{\phcb\phc}$ is specified as follows
\beqs\beq M_{\phcb\phc}=\frac{pN}{2\ft^{p+2}}\bcs
\mtta{\ft+\fp}{(p+1)\sg^2}{(p+1)\sg^2}{\ft+\fp}&
\hspace*{-1.6mm}\mbox{for $K=\tkas$},\\ \hspace*{3mm}2\fp\
\diag\lf1,1\rg & \hspace*{-1.6mm}\mbox{for $K=\tkbs$,}
\ecs\label{Mfs} \eeq
where $\ft=1-\sg^2$ and $\fp=1+p\sg^2$. Upon diagonalization of
$M_{\phcb\phc}$ for $K=\tkas$ we find its eigenvalues which are
\bea
\label{kpm}\kp_+=\kp=pN\fp/\ft^{p+2}~~\mbox{and}~~\kp_-=pN/\ft^{p+1},\eea\eeqs
from which $\kp_+$ equals to the double eigenvalue $\kp$ of
$M_{\phcb\phc}$ for $K=\tkbs$. Inserting \eqs{hpar}{kab} in the
second term of the right-hand side of \Eref{action1} we can define
the canonically normalized fields which are denoted by hat.
Combining \eqs{Mfs}{kpm} we can establish for both $\tkis$ in
\Eref{tkis} a unique expression regarding the normalized inflaton
$\se$, i.e.,
\beq \label{Je}
{d\se}/{d\sg}=J=\sqrt{2\kp}~~\mbox{for}~~\sg<1.\eeq
This result is identical with the $\sg-\se$ relation obtained for
$K=K_{1\rm T}$ (or $\wtilde K_{1\rm T}$) in \cref{actpole}. For
the remaining fields of the $\phcb-\phc$ system we find
\bea &&\hspace*{-0.3cm}  \widehat{\theta}_\pm
=\sqrt{\kp_\pm}\sg\theta_\pm,\>\>\widehat \theta_\Phi =
\sqrt{2\kp_-}\sg\lf\theta_\Phi-{\pi}/{4}\rg
~\mbox{for $K=\tkas$},\nonumber \\
&&\hspace*{-0.3cm} \widehat{\theta}_\pm
={\sqrt{\kp}}\sg\theta_\pm,\hspace*{0.4cm}\widehat \theta_\Phi =
\sqrt{2\kp}\sg\lf\theta_\Phi-{\pi}/{4}\rg\hspace*{0.4cm}\mbox{for
$K=\tkbs$,}\nonumber \eea
where $\th_{\pm}=\lf\bar\th\pm\th\rg/\sqrt{2}$ -- needless to say,
$(s,\bar s)$ are canonically normalized since $\vevi{K_{SS^*}}=1$.
Note, in passing, that the spinors $\psi_S$ and $\psi_{\Phi\pm}$
associated with the superfields $S$ and $\Phi-\bar\Phi$ are
normalized similarly, i.e.,
$\what\psi_{\Phi+}=\sqrt{\kp}\psi_{\Phi+}$ with
$\psi_{\Phi\pm}=(\psi_\Phi\pm\psi_{\bar\Phi})/\sqrt{2}$.

\renewcommand{\arraystretch}{1.4}
\begin{table}[!t]
\caption{\normalfont Mass-squared spectrum for $K=\tkis$
($\ell=1,2$) along the path in \Eref{inftr}.}
\begin{tabular}{c|c|c|c|c}\toprule
%
{\sc Fields}&{\sc Eigen} & \multicolumn{3}{c}{\sc Masses
Squared}\\\cline{3-5}
&{\sc states}&&{$K=\tkas$}&{$K=\tkbs$}\\\colrule
4 real &$\widehat\theta_{+}$& $
m_{\theta+}^2$&\multicolumn{2}{|c}{$6\Hhi^2$}\\\cline{4-5}
scalars&$\widehat \theta_\Phi$ & $ m_{
\theta_\Phi}^2$&\multicolumn{2}{|c}{$M^2_{BL}+6\Hhi^2$}\\\cline{4-5}
&$s, \bar{s}$ & $
m_S^2$&\multicolumn{2}{c}{$6\lf1/\nst+2\ft^{p+2}/pN\sg^2\fp\rg\Hhi^2$}\\\hline
1 gauge & \multirow{2}{0.25in}{$A_{BL}$} &
\multirow{2}{0.25in}{$M_{BL}^2$}&\multirow{2}{0.8in}{$2g^2pN\sg^2/\ft^{p+1}$}
&\multirow{2}{0.9in}{$2g^2pN\sg^2\fp/\ft^{p+2}$}\\
boson& & &\multicolumn{1}{c|}{}&\\
\colrule
$4$ Weyl & $\what \psi_\pm$ &  $m^2_{ \psi\pm}$ &
\multicolumn{2}{c}{${6\ft^{p+2}\Hhi^2/pN\sg^2\fp}$}
\\\cline{4-5} spinors &$\ldu_{BL}, \widehat\psi_{\Phi-}$&
$M_{BL}^2$&{$2g^2pN\sg^2/\ft^{p+1}$}
&{$2g^2pN\sg^2\fp/\ft^{p+2}$}\\\botrule
\end{tabular}\label{tab1}
\end{table}
%

We can verify that the inflationary direction in \Eref{inftr} is
stable w.r.t the fluctuations of the fields
$\chi^\al=\theta_+,\theta_\Phi$ and $S$. To this end, we construct
the mass-squared spectrum of those scalars. Taking the limit
$M\ll1$ we find the expressions of the masses squared
$m^2_{\chi^\al}$ arranged in \Tref{tab1}, which approach rather
well the exact expressions taken into account in our numerical
computation. These expressions assist us to appreciate the role of
$\nst>0$ in retaining positive $m^2_S$. Moreover,
$m^2_{\chi^\al}\gg\Hhi^2=\Vhio/3$ for $\sgf\leq\sg\leq\sgx$ --
where $\sgx$ and $\sgf$ are the values of $\sg$ when
$\ks=0.05/{\rm Mpc}$ crosses the horizon of \fhi\ and at the end
of \fhi\ correspondingly. In \Tref{tab1} we display also the
masses, $M_{BL}$, of the gauge boson $A_{BL}$ and the masses of
the corresponding fermions with the eigenstate $\what \psi_\pm$
defined as $\what \psi_\pm =(\what{\psi}_{\Phi+}\pm
{\psi}_{S})/\sqrt{2}$. Since $\Gbl$ is broken during \fhi, no
cosmic strings are formed after its termination.

The derived mass spectrum can be employed in order to find the
one-loop radiative corrections, $\dV$, to $\Vhi$. Considering
SUGRA as an effective theory with cutoff scale equal to $\mP$, the
well-known Coleman-Weinberg formula can be employed
self-consistently taking into account only the masses which lie
well below $\mP$, i.e., all the masses arranged in \Tref{tab1}
besides $M_{BL}$ and $m_{\th_\Phi}$ -- cf. \cref{nmBL, jhep}. The
resulting $\dV$ lets intact our inflationary outputs, provided
that the renormalization-group mass scale $\Lambda$, is determined
by requiring $\dV(\sgx)=0$ or $\dV(\sgf)=0$. E.g., imposing the
former condition for $\nst=1$ we find
$\Ld\simeq(2.2-5.1)\cdot10^{-5}$ for the allowed parameter space
exposed below. Under these circumstances, our inflationary
predictions can be exclusively reproduced by using $\Vhi$ in
\Eref{Vhio}.

\section{Inflationary Requirements} \label{res}

Applying the standard formulas quoted in \cref{actpole} for
$V_{\rm I}=\Vhi$, we can compute a number of observational
quantities, which assist us to qualify our inflationary setting.
Namely, we extract the number $\Ns$ of e-foldings that the scale
$\ks$ experiences during \fhi\ and the amplitude $\As$ of the
power spectrum of the curvature perturbations generated by $\sg$
for $\sg=\sgx$. These observables must be compatible with the
requirements \cite{actin}
\begin{align} \label{prob}
&\Ns\simeq61.3+\frac{1-3\wrh}{12(1+\wrh)}\ln\frac{\pi^2g_{\rm rh*}\Trh^4}{30\Vhi(\sgf)}+\nonumber\\
&\frac14\ln{\Vhi(\sgx)^2\over g_{\rm
rh*}^{1/3}\Vhi(\sgf)}~~\mbox{and}~~\As^{\frac12}\simeq4.618\cdot10^{-5},\end{align}
where we take into account that \fhi\ is followed in turn by an
oscillatory phase with mean equation-of-state parameter $w_{\rm
rh}$, radiation and matter domination. We estimate $\wrh$ as
detailed in Appendix A and we find $\wrh\simeq(0.12-0.3)$ in the
major part of the parameter space -- note that $\wrh$ decreases
with $N$.

\begin{figure}[!t]
\includegraphics[width=60mm,angle=-90]{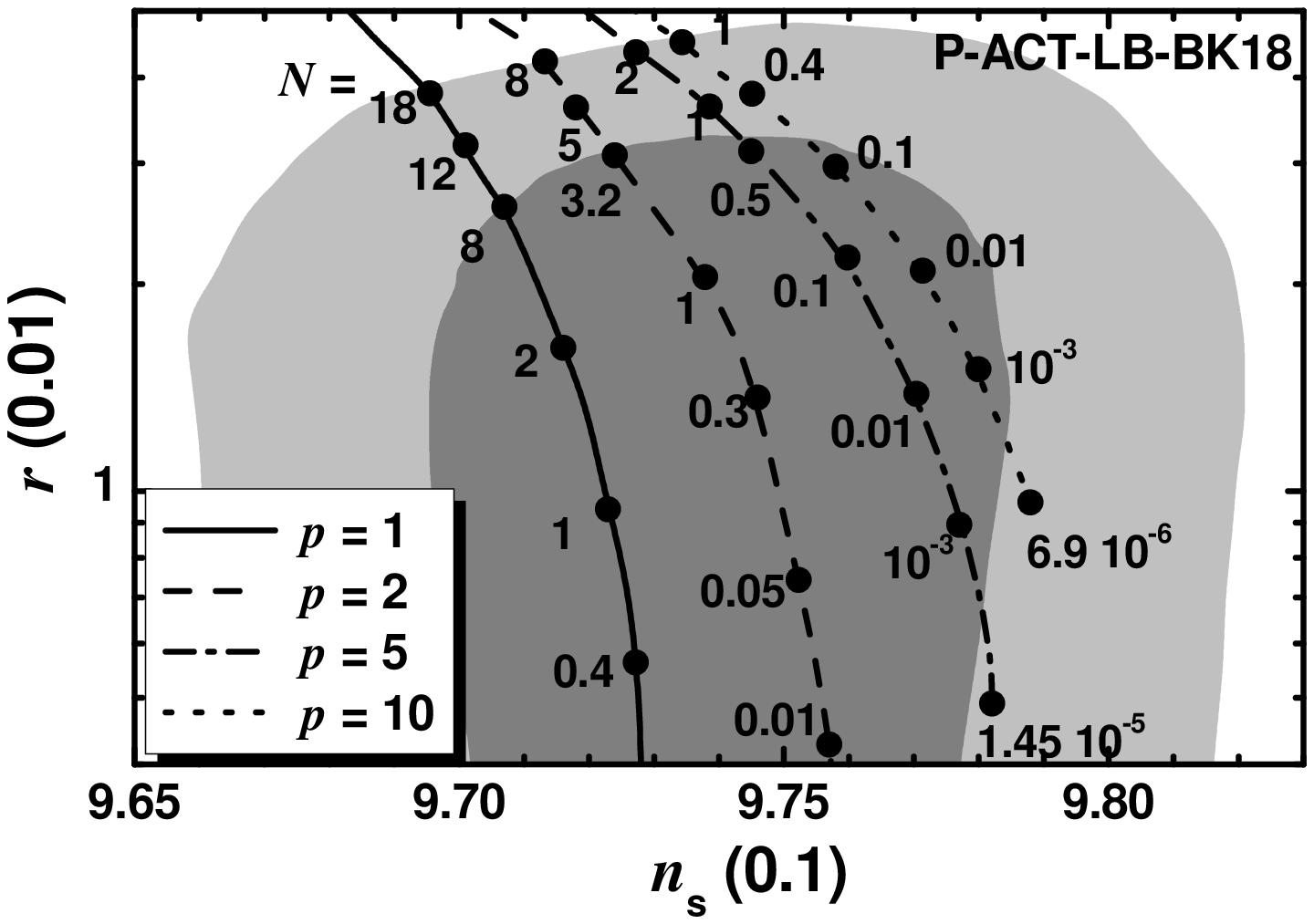}
\caption{\sl Allowed curves in the $\ns-r$ plane for various $p$'s
shown in the legend and varying $N$ as shown along the curves. The
marginalized joint $68\%$ [$95\%$] regions from \actcf\ data are
depicted by the dark [light] shaded contours.}\label{fig1}
\end{figure}

The constraints above allows us to extract $\ld$ and $\sgx$ for
any given $p$, $N$ and $M$. However, $M$ is irrelevant for the
inflationary observables, provided that it is confined to values
much lower than $\mP$. Motivated by the unification of the gauge
coupling constants with the particle content of MSSM, we determine
$M$ identifying the unification scale
$\Mgut\simeq2/2.433\cdot10^{-2}$ with the value of $M_{BL}$ -- see
\Tref{tab1} -- at the SUSY vacuum in \Eref{vevs}. Given that
$\vev{\ft}\simeq1$ for $M\ll1$, we obtain, for both $\tkis$ in
\Eref{tkis}
\beq \label{Mg} M\simeq{\Mgut}/{g\sqrt{2Np}}\eeq
with $g\simeq0.7$ being the value of the GUT gauge coupling
constant. The success of our inflationary setting is assured if
\beq \label{Nmin}
M\lesssim\mP~~\Rightarrow~N\gtrsim\nmin={\Mgut^2}/{2g^2p\mP^2}\eeq
and so a lower bound on $N$ is applied. We are then left with only
two free input parameters $(p,N)$ which can be constrained
computing $\ns$, its running, $\as$ and $r$. These quantities can
be found approximately applying the analytic expressions displayed
in \cref{actpole}. Namely, we found
\beqs\beq \label{ns} \ns\simeq 1-\frac{p+2}{(p+1)\Ns},~~
\as\simeq-\frac{p+2}{(p+1)N^2_\star}\eeq
which are independent of the exponent $n=4$ of $\sg$ in
\Eref{Vhio} whereas $r$ for the specific $n$ value is written as
\beq \label{rpN} r\simeq
{4(4^ppN)^{1/(p+1)}}/{(p+1)N_\star^{(p+2)/(p+1)}}. \eeq\eeqs
The expressions above for $\ns$ and $\as$ are independent from $N$
in the lowest order expansion in powers of $1/\Ns$. This feature,
however, is not totally kept as we show numerically in
\fref{fig1}, where we compare our outputs against the \actc\ data
\cite{actin} in the $\ns-r$ plane. We draw solid, dashed,
dot-dashed and dotted lines for $p=1, 2, 5$ and $10$ respectively
and show the variation of $N$ along each line. For the two latter
$p$'s the corresponding \nmin's from \Eref{Nmin} are also
displayed. We clearly see that $\ns$ increases with $p$ and $r$
with $N$ and so the whole shaded region favored by \actc\ is
covered varying $p$ and $N$. As a bonus, the resulting $r$'s for
$(p,N)$'s of order unity lie within the reach of planned
experiments \cite{det} aiming to discover primordial gravitational
waves.


\begin{figure}[!t]
\includegraphics[width=60mm,angle=-90]{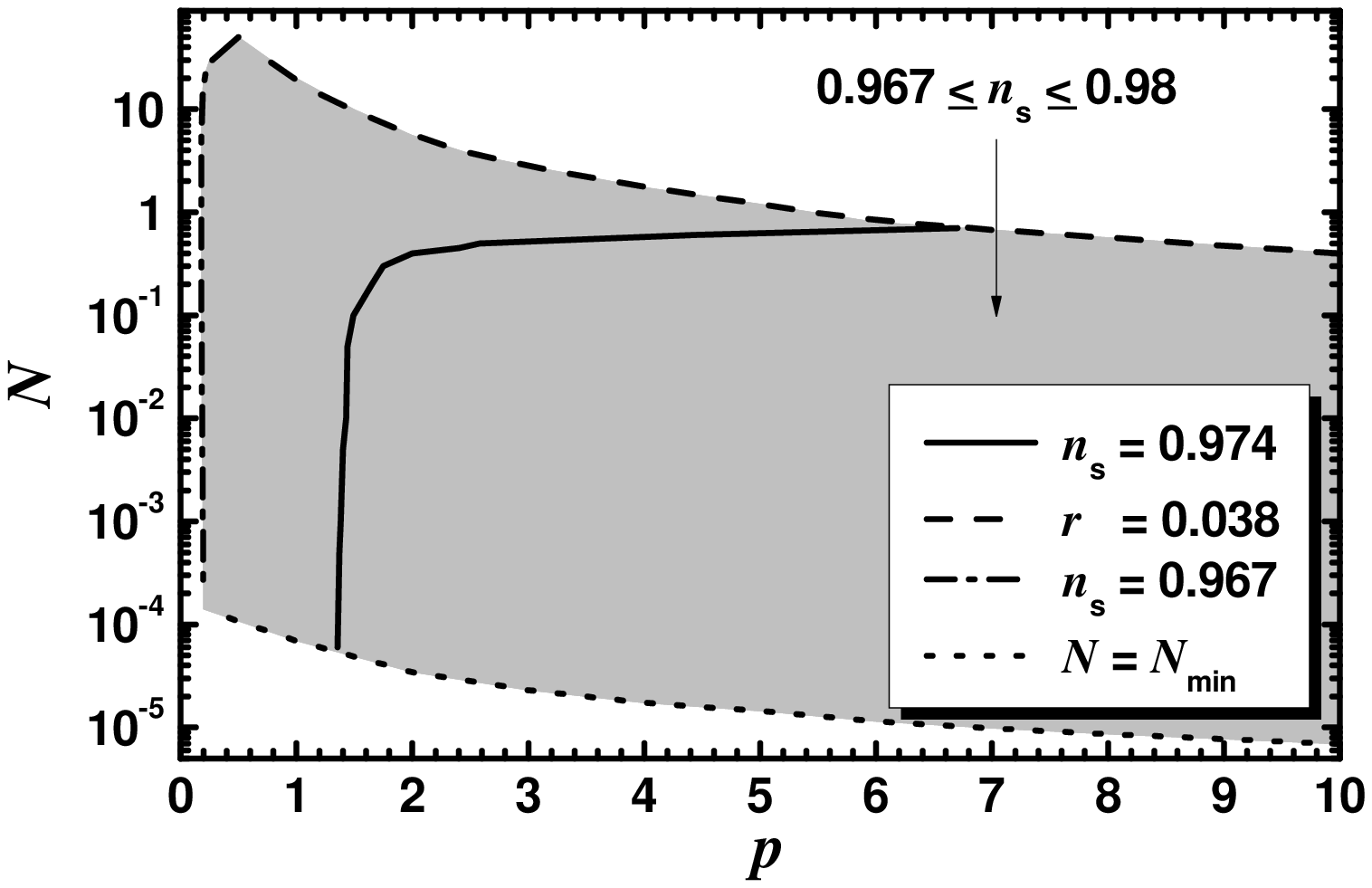}\vspace*{-0.15cm}
\hfill \caption{\sl\small  Allowed (shaded) region as determined
by \eqss{data}{prob}{Nmin} in the $p-N$ plane. Along the solid
line we fix $\ns=0.974$. The conventions adopted for the various
lines are also shown.}\label{fig2}
\end{figure}

The variation of the observables in \Fref{fig1} reveals that our
free parameters can be constrained in the $p-N$ plane. The
relevant allowed region is shown in \Fref{fig2}. It is bounded by
{\ftn\sf (i)} the dashed line which originates from the bound on
$r$ in \Eref{data}, {\ftn\sf  (ii)} the dot-dashed line which
comes from the lower bound on $\ns$ in \Eref{data} and {\ftn\sf
(iii)} the dotted line along which \Eref{Nmin} is saturated --  we
take by hand a maximal $p=10$ since no upper bound on $p$ can be
inferred from the corresponding bound on $\ns$ in \Eref{data}.
Fixing $\ns$ to its central value in \Eref{data}, we obtain the
solid line along which we obtain
\beq \label{res1} 1.355\lesssim
p\lesssim6.7~~~\mbox{and}~~~6\cdot10^{-5}\lesssim N\lesssim0.7\eeq
with $\Ns\simeq(54.7-56.8)$, $\as\simeq -4.7\cdot10^{-4}$ and
$r\gtrsim 2.8\cdot10^{-4}$. The obtained $|\as|$'s remain
negligibly small being, thereby, consistent with its $95\%$ c.l.
allowed margin in \cref{actin}.

Note, finally, that our proposal is stabilized against corrections
from higher order terms -- e.g., $(\bar\Phi\Phi)^l$ with $l>1$ in
\Eref{whi} -- since $\sgx<1$, as dictated by \Eref{Je}. The closer
to unity $\sgx$ is chosen, the largest $\Ns$ is obtained. To
quantify the relevant tuning we estimate $\Dex=1-\sgx$ which
ranges in the interval $(0.8-28)\%$ for the values in \Eref{res1}.
The maximal $\Dex$ values are obtained for the largest $p$'s and
$N$'s indicating, thereby that naturalness prefers $N$ and $p$
values of order unity.

\section{Post-Inflationary Completion}\label{post}

Our inflationary setting can be easily embedded in a $B-L$
extension of MSSM promoting to gauge the pre-existing global
$U(1)_{B-L}$ -- cf. \cref{nmBL, tmhi}. The terms of the total
super- and \Kaa potentials which control the coexistence of the
inflationary and the MSSM sectors are
\beqs\bea \label{dW} \dW&= &\ld_{\mu} S\hu\hd\ +\  h_{ijN}
\sni L_j \hu +\lrh[i]\phcb N^{c2}_i; \hspace*{0.7cm}\\
\dK&=&\nst\ln\lf1+\mbox{$\sum_{a=1}^{4}$}|X_\al|^2/\nst\rg\label{dK}\eea\eeqs
with~~$X_\al=\tilde N^{c}_i, \tilde L_j,\hu$ and $\hd$ where
$\ssni$ and $\tilde L_j$ are the scalar superpartners of the
right-handed neutrinos $\sni$ and left-handed leptons $L_j$
respectively. Here we adopt the notation and the $B-L$ and $R$
charges of the various superfields as displayed in Table 1 of
\cref{nmBL}. For simplicity we use the same prefactors of $\ln$ in
all terms of $\dK$.

%
%

We assume that $X_\al$ are stabilized at zero during \fhi\ and so
the inflationary trajectory in \Eref{inftr} has to be supplemented
by the condition
\beq\label{inftr1} \vevi{\hu}=\vevi{\hd}=\vevi{\ssni}=\vevi{\tilde
L_j}=0.\eeq
The stability of this path can be checked, parameterizing the
complex fields above as we do for $S$ in \Eref{hpar} -- note that
we use small letters for the real and imaginary components. The
relevant masses squared are listed in \Tref{tab2} where we see
that $m_{i\tilde \nu^c}^2>0$, $m_{i\tilde l}^2>0$ and $m^2_{h+}>0$
for any $\sgf<\sg<1$ -- here $m_{h\pm}$ are the eigenvalues
associated with the Higgs eigenstates
\beq h_\pm=(h_u\pm{h_d})/\sqrt{2}\>\>\>\mbox{and}\>\>\> {\bar
h}_\pm=({\bar h}_u\pm{\bar h}_d)/\sqrt{2}. \eeq
On the other hand, the positivity of $m^2_{h-}$ dictates the
establishment of the inequality -- cf. \cref{tmhi}:
\beq \lm<\ld(1+\nst)\sgf^2/2\nst.\label{lmb} \eeq
We can verify numerically that the requirement above is fulfilled
for $\nst=1$ and $\lm\lesssim2\cdot10^{-5}$. Despite this low
upper bound, $\lm$ can become compatible with $\mmgr\sim1$ as we
show in \Sref{secmu}, where the generation of the $\mu$ term of
MSSM is analyzed. On the other hand, the implementation of nTL is
explored in \Sref{seclepto} and our final results, including the
post-inflationary constraints, are given in \Sref{respost}. Note
in passing that due to the extra contributions of \Tref{tab2}
$\dV$ increases and so $\Ld$ should be readjusted. E.g., for $p=2$
we find $\Ld\simeq(4-5)\cdot10^{15}~\GeV$  -- hereafter we restore
units, i.e., we take $\mP=2.43\cdot10^{18}~\GeV$.

\renewcommand{\arraystretch}{1.4}
\begin{table}[!t]
\caption{\normalfont Mass-squared spectrum of the non-inflaton
sector along the path in \eqs{inftr}{inftr1} for $K=\tkis+\dK$
with $\ell=1,2$.}
\begin{tabular}{c|c|c|c|c}\toprule
{\sc Fields}&{\sc Eigen-} & \multicolumn{3}{c}{\sc Masses
Squared}\\
&{\sc states}&\multicolumn{3}{c}{}\\ \colrule
26 Real & $h_{\pm},{\bar h}_{\pm}$ &
$m_{h\pm}^2$&\multicolumn{2}{c}{$3\Hhi^2\lf1+1/\nst\pm{2\lm}/{\ld\sg^2}\rg$}\\\cline{4-5}
Scalars  & $\tilde \nu^c_{i}, \bar{\tilde\nu}^c_{i}$ & $
m_{i\tilde
\nu^c}^2$&\multicolumn{2}{c}{$3\Hhi^2\lf1+1/\nst+8\ld^2_{iN^c
}/\ld^2\sg^2\rg$} \\\cline{4-5}
& $\tilde l_{i}, \bar{\tilde l}_{i}$ & $ m_{i\tilde
l}^2$&\multicolumn{2}{c}{$3\Hhi^2(1+1/\nst)$} \\\colrule
$3$ Weyl&\multirow{2}{0.15in}{$N_i^c$}& \multirow{2}{0.27in}{$
M_{{iN^c}}^2$}&\multicolumn{2}{c}{\multirow{2}{0.9in}{$24\ld^2_{iN^c}\Hhi^2/\ld^2\sg^2$}}\\
Spinors &&&\multicolumn{2}{c}{}\\\botrule
\end{tabular}\label{tab2}
\end{table}
\renewcommand{\arraystretch}{1.}

\subsection{\sf\small Generation of the $\mu$ Term of MSSM} \label{secmu}

The origin of the $\mu$ term of MSSM can be explained if we
combine the terms of \Eref{whi} with the first term in \Eref{dW}
working at the level of SUSY. In fact, the total low energy
potential is
\beq V_{\rm tot}=V_{\rm F}+V_{\rm soft},\label{vtot}\eeq
where the first term includes the SUSY F-term contributions into
$V_{\rm tot}$ which read
\beq\begin{aligned}  V_{\rm F}&=& K_{\rm SUSY}^{\al\bbet}
(W+\dW)_{\al} (W+\dW)^*_{\bbet}~~~~~~~~~~~~~~~~~~~~~~~~~~~~~
\\&=&\ld^2|\phcb\phc-M^2/2|^2+\ld^2|S|^2\lf|\phcb|^2+|\phc|^2\rg/Np+\cdots\label{vsusy}
\end{aligned}\eeq
where $K_{\rm SUSY}$ can be obtained by expanding $\tkis+\dK$ --
see \eqs{tkis}{dK} -- for low field values with result
\beq K_{\rm
SUSY}=Np\lf|\phc|^2+|\phcb|^2-\phcb\phc-\phcb^*\phc^*\rg+|S|^2+|X^\al|^2.\eeq
Also ellipsis in \Eref{vsusy} includes terms which vanish at the
SUSY vacuum due to the zero v.e.vs of $\hu, \hd$ and $\ssni$. On
the other hand, the contribution into $V_{\rm tot}$ from the soft
SUSY-breaking terms read
\beq V_{\rm soft}= \lf\ld A_\ld S \phcb\phc- \ld \aS M^2S/2 + {\rm
h. c.}\rg+ m_{\gamma}^2\left|X^\gamma\right|^2, \label{Vsoft} \eeq
where $X^\gamma=S, \Phi, \bar\Phi, \hu$ and $\hd$. Also
$m_{\gamma}, A_\ld$ and $\aS$ are soft SUSY-breaking mass
parameters much lower than $M$. Rotating $S$ in the real axis by
an appropriate $R$ transformation, choosing conveniently the
phases of $\Ald$ and $\aS$ -- so as $V_{\rm tot}$ in \Eref{vtot}
to be minimized -- and substituting the $\phc$ and $\phcb$ values
from \Eref{vevs}, we get
\beqs\beq \vev{V_{\rm tot}(S)}= \ld^2{M^2S^2}/{Np} -2\ld\am\mgr
M^2S, \label{Vol} \eeq
where $\mgr$ is the $\Gr$ (gravitino) mass and
\beq \am={(|A_\ld| + |{\rm a}_{S}|)}/{2\mgr}>0\eeq
is a parameter of order unity which parameterizes our ignorance
about the dependence of $|A_\ld|$ and $|{\rm a}_{S}|$ on $\mgr$.
The extermination condition of $\vev{V_{\rm tot}(S)}$ w.r.t $S$
shifts slightly $\vev{S}$ from zero in \Eref{vevs} --
cf.~\cref{dvali} -- as follows
\beq \label{vevS}{d}\vev{V_{\rm tot}(S)}/{d S}
=0~~\Rightarrow~~\vev{S}= 2Np\ \am\mgr/{\ld}.\eeq\eeqs
The generated $\mu$ parameter from the first term in \Eref{dW},
\beq \label{mu} \mu =\lm \vev{S}= 2\lm Np\ \am\mgr/{\ld},\eeq
can be comparable to $\mgr$ for $N\sim1$, if $\lm$ is of the same
order of magnitude with $\ld$. Since $\ld\sim 10^{-6}$, due to
$\ld-\As$ condition implied by the rightmost expression in
\Eref{prob} -- see \cref{actpole} --, the resulting $\lm$ can
become compatible with \Eref{lmb} assuring, thereby, the stability
of the path in \Eref{inftr1} during \fhi.

\subsection{\sf\small Non-Thermal Leptogenesis and $\Gr$ Constraint}\label{seclepto}

Besides the generation of the $\mu$ term, $\dW$ in \Eref{dW}
allows for the implementation of nTL via the direct decay of the
inflaton into $\rhni$. For the success of this process, however,
one has to restrict the decay channel of the inflaton into $\hu$
and $\hd$ which opens via the first term of $\dW$. This implies
constraints on $\lm$ or, via \Eref{mu}, on the ratio $\mmgr$.

More explicitly, when \fhi is over, the (canonically normalized)
inflaton
\beq\dphi=\vev{J}\dph\>\>\>\mbox{with}\>\>\> \dph=\phi-M
\>\>\>\mbox{and}\>\>\>\vev{J}\simeq\sqrt{2Np}\label{dphi} \eeq
continues to roll down towards the SUSY vacuum, \Eref{vevs}, where
it acquires mass -- reducing with increasing $N$ and $p$ -- given
by
\beq \label{msn} %
\msn=\left\langle\Ve_{\rm HI,\se\se}\right\rangle^{1/2}\simeq{\ld
M}/{\sqrt{Np}}.\eeq
After a brief stage of tachyonic preheating \cite{preheating}
which does not lead to significant particle production
\cite{garcia}, \dphi\ settles into a phase of damped oscillations
initially around zero -- where $\Vhi$ has a maximum -- and then
around one of the minima of $\Vhi$ similarly to the picture
described in Appendix B of \cref{hi5} for the non-minimal HI.
Whenever \dphi\ passes through zero, several particles (mostly
$A_{BL}$ bosons) may be produced via the mechanism of instant
preheating \cite{instant}. This process becomes more efficient as
$N$ decreases since the amplitude of the first oscillation
\beq \sgm \simeq \sqrt{6/(16 + 3Np)}\mP \label{sgm} \eeq
increases, increasing thereby the frequency of the violation of
the adiabaticity criterion \cite{instant, garcia}. Given that the
efficiency of this mechanism requires a rather large (let say
$50-70$) number of passages of $\dphi$ through zero, we
intuitively expect that our estimates below are more accurate for
$N>0.01$ which ensures a less frequent passage of \dphi\ through
zero weakening, thereby, the effects from preheating.


Nonetheless the standard perturbative approach to the $\dphi$
decay provides a very efficient total decay rate
\beq \Gsn=\GNsn+\Ghsn, \label{Gtot}\eeq
where the individual decay widths are
\beqs\bea \label{Ga}
\GNsn&=&({g_{iN^c}^2}\msn/{16\pi})\lf1-{4\mrh[
i]^2}/{\msn^2}\rg^{3/2}; \hspace*{0.5cm}\\
\Ghsn&=&{2}g_{H}^2\msn/{8\pi} \label{Gb}\eea
with the relevant coupling constants being
\beq
g_{iN^c}=\sqrt{2}\ld_{iN^c}/{\vev{J}}\>\>\>\>\mbox{and}\>\>\>\>
g_{H}={\lm}/{\sqrt{2}}.\label{gs}\eeq\eeqs
Moreover, $\mrh[i]$ are the Majorana masses of $\rhni$ identified
by the third term in \Eref{dW} as
\beq\label{mrhi} \mrh[i]=\sqrt{2}\ld_{iN^c}M~~\mbox{with}~~
\ld_{iN^c}\leq4\pi\eeq
to be consistent with the perturbative bound. The lagrangian terms
which describe the decay channels above are similar to those given
e.g. in \cref{nmBL}. Note that three-particle decay of $\dphi$ is
suppressed as noted there and possible perturbative decay into
$A_{BL}$ bosons is kinetically blocked. Thanks to \Gsn\ the
universe is reheated at a temperature \cite{reh,reh1}
\beq\Trh= \left({90\over\pi^2g_{\rm
rh*}}\right)^{1/4}\lf{2(\wrh+1)\over
5-3\wrh}\mP\Gsn\rg^{1/2}\label{Trh}\eeq
with $g_{\rm rh*}=228.75$ the MSSM effective number of
relativistic degrees of freedom.

For $\mrh[i]\gg\Trh$, the out-of-equilibrium decay of $\rhni$ via
the second term in \Eref{dW} generates a $L$-asymmetry yield which
is partially converted through sphaleron effects into a yield of
the observed $B$ asymmetry of the universe \cite{lept, zhang}
\beq Y_B=-0.35\cdot2\cdot{5-3\wrh\over4(1-\wrh)}{\Trh\over\msn}
{\GNsn\over\Gsn}\ve_L.\label{yb}\eeq
The simplified formula above can be derived assuming normal
hierarchy for $\mrh[i]$ -- i.e., $\mrh[1]\ll \mrh[2],\mrh[3]$ --
and that $\dphi$ decays via \GNsn\ in \Eref{Ga} exclusively into
$\rhna$. Under these circumstances, we can obtain a maximal value
for the $L$ asymmetry $\ve_L$ which is \cite{sasa, dreeslept}
\beq\label{el} \ve_L = -\frac {3}{8\pi}\frac{\mntau
\mrh[1]}{\vev{\hu}^2} \>\>\mbox{where}\>\> \mntau=\sqrt{\Delta
m^2_\oplus}\simeq0.05\>{\rm eV} \eeq
is the mass of heaviest light neutrino $\nu_\tau$ and $\Delta
m^2_\oplus$ \cite{neutop} the atmospheric neutrino mass-squared
difference. Also, we set $\vev{\hu}=174~\GeV$ adopting the large
$\tan\beta$ regime of MSSM. The validity of \Eref{yb} requires the
kinematic allowance of the decay $\dphi\to2\rhna$ and the
avoidance of any erasure of the produced lepton asymmetry yield.
These conditions can be fulfilled if we respectively demand
\beq\label{kin}
\msn\geq2\mrh[1]\>\>\>\mbox{and}\>\>\>\mrh[1]\gtrsim 10\Trh.\eeq
The result of \Eref{yb} has to meet the observational one
\cite{act}
\beq \Yb=\lf8.75\pm0.085\rg\cdot10^{-11}~~\mbox{at $95\%$
c.l.}\label{ybdat}\eeq

On the other hand, $\Trh$  must be compatible with constraints on
the $\Gr$ abundance, $Y_{3/2}$, at the onset of \emph{Big Bang
nucleosynthesis} ({\sf\small BBN}), which is estimated to be
\cite{kohri}
\beq\label{ygr} \Yg\simeq
2.9\cdot10^{-13}(\Trh/\EeV)/(3-\wrh),\eeq
where we take into account only thermal production of $\Gr$ and
assume that $\Gr$ is much heavier than the MSSM gauginos. On the
other hand, $\Yg$  is bounded from above in order to avoid
spoiling the success of the BBN. For the typical case where $\Gr$
decays with a tiny hadronic branching ratio, the bounds on $\Yg$
imply the following upper bounds on $\Trh$ \cite{brand,kohri}
\beq  \label{ygdat} \Trh\lesssim\left\{\bem
%
0.53~\EeV\hfill \cr
5.3~\EeV\hfill \cr\eem
\right.\>\>\>\>\mbox{for}\>\>\>\>\mgr\simeq\left\{\bem
10.6~\TeV\,,\hfill \cr
13.6~\TeV\,,\hfill \cr \eem
\right.\eeq
at $95\%$ c.l. The bounds above can be somehow relaxed in the case
of a stable $\Gr$ -- cf. \cref{fhi5}.

\begin{figure}[!t]
\includegraphics[width=60mm,angle=-90]{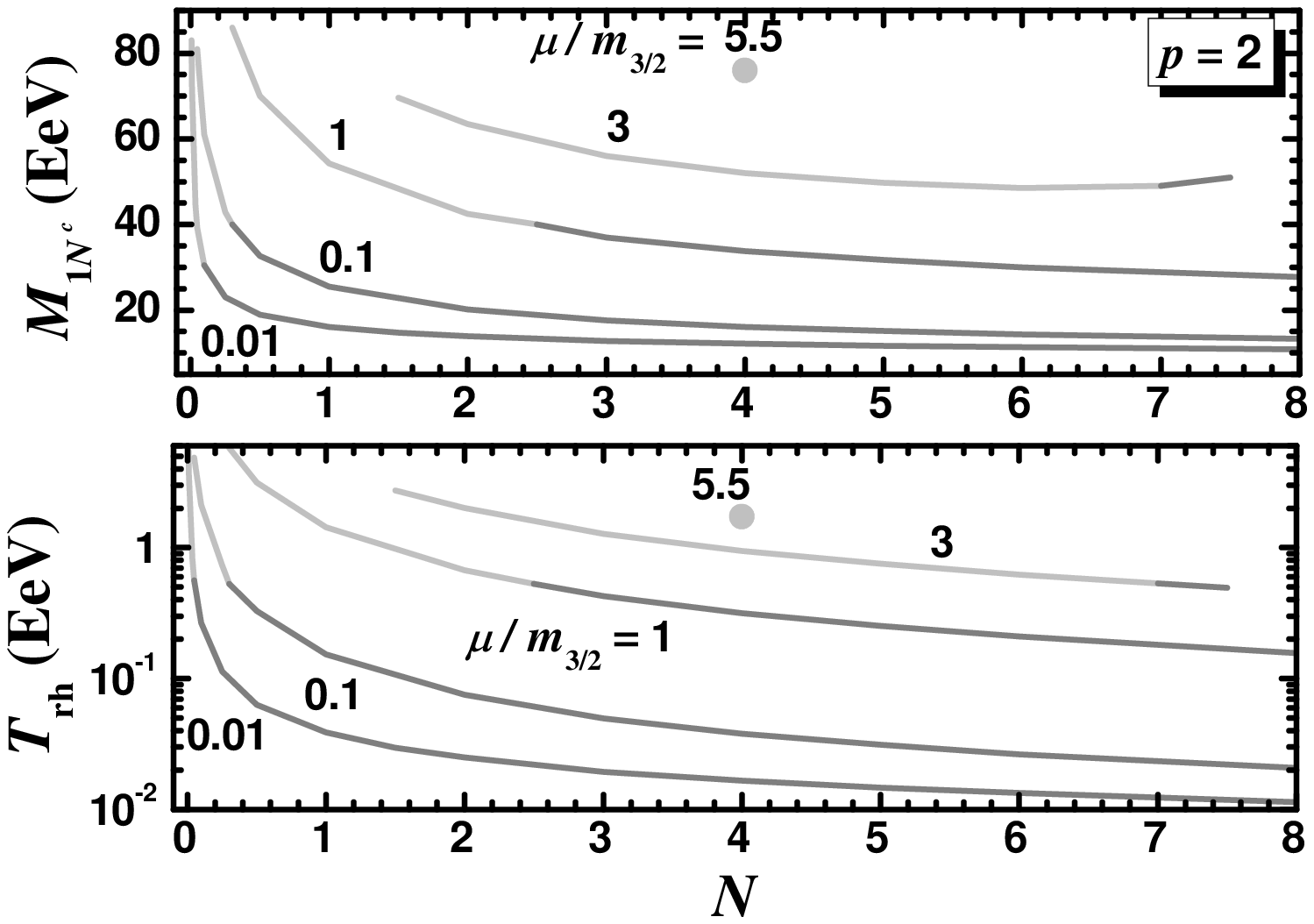}
\hfill\caption{\sl\small  Contours yielding the central $Y_B$ in
\Eref{ybdat} consistently with
\eqssss{data}{prob}{lmb}{kin}{ygdat} in the $N-\mrha$ and $N-\Trh$
planes for $p=2$, $\nst=1/3$ and various $\mmgr$ values indicated
on the curves -- recall that $1~\EeV=10^{9}~\GeV$. The gray [light
gray] segments of the lines are allowed by \Eref{ygdat} for
$\mgr=10.6~\TeV$ [$\mgr=13.6~\TeV$].}\label{frhn}
\end{figure}

\subsection{\sf\small  Results}\label{respost}


The compatibility of the restrictions in
\eqsss{lmb}{kin}{ybdat}{ygdat} together with those from
\eqs{data}{prob} is illustrated in \Fref{frhn} where we present
allowed curves in $N-\mrha$ and $N-\Trh$ planes for $p=2$,
$\nst=1/3$ and $\mmgr=0.01, 0.1, 1$ and $3$. The chosen $p$ value
is very close to the $p$ values which yields the central $\ns$ in
\Eref{data}, as shown in \Fref{fig2}, and the one for $\nst$
assists us to overcome the constraint of \Eref{lmb} without
extensive tuning. Moreover, the considered $\mmgr$ ratios are
encountered in various recent studies \cite{klp, gabit, gmssm} of
the parameter space of MSSM with $\mgr$ values comparable to those
considered in \Eref{ygdat}. Despite the obtained $\mu$ value is
quite low for $\mmgr=0.01$, it is in general still allowed by the
LEP \cite{lep} exclusion limit on the chargino mass.

The ratio $\mmgr$ controls the $\lm$ value derived by \Eref{mu}
and via it the contribution of $\Ghsn$ in \Eref{Gb} into the
$\Gsn$ -- see \Eref{Gtot}. As expected, the required by
\Eref{ybdat} $\mrha$ increases with $\mmgr$ in order to reinforce
the contribution of $\GNsn$ -- see \Eref{Ga} -- into $\Gsn$. The
increase of $\mmgr$ is constrained directly by \Eref{lmb} and
indirectly by \Eref{ygdat} since it causes an increase of $\Trh$
-- see \Eref{Trh}. The gray [light gray] segments of the lines
displayed in \Fref{frhn} are allowed by \Eref{ygdat} for
$\mgr=10.6~\TeV$ [$\mgr=13.6~\TeV$]. As shown from the lower plot
in \Fref{frhn} the lines corresponding to $\mmgr=0.01$ and $0.1$
terminate for low $N$ values due to the upper bound on $\Trh$ in
\Eref{ygdat}. On the other hand, for $\mmgr=1$ and $3$ this
termination occurs due to the saturation of the bound on $\lm$ in
\Eref{lmb}. The $\mmgr=3$ line is cut for $N=7.5$ due to the
leftmost inequality in \Eref{kin}.  The competition of these two
aforementioned constraints causes a progressive reduction of the
length of the allowed contours as $\mmgr$ increases beyond $3$
until they shrink to one point for $\mmgr\simeq5.5$ corresponding
to $N=4$, $\mrha=76~\EeV$ and $\Trh=1.73~\EeV$. Along the curves
of \Fref{frhn} we obtain
\beq 0.035\lesssim \lm/10^{-6}\lesssim40~~\mbox{and}~~0.14\lesssim
\msn/\ZeV \lesssim32,\label{resp}\eeq
where $1~\ZeV=10^{12}~\GeV$. The largest $\lm$ is obtained for
$\mmgr=1$ and $N=0.3$ whereas the lowest one for $\mmgr=0.01$ and
$N=8$. On the other hand, $\msn$ decreases with increasing $N$
independently from $\mmgr$. Had we used larger $\nst$ values most
of the curves with $\mmgr>1$ would have been excluded. Therefore,
our scheme primarily prefers the hierarchy $\mu\leq\mgr$ which is
defended by various analysis \cite{klp, gabit, gmssm} of MSSM
parameter space.

As a bottom line, nTL is a realistic possibility within our
setting since it can be reconciled with the $\Gr$ constraint for
$\mgr\sim10~\TeV$ and several MSSM versions.



\section{Conclusions}\label{con}

Prompted by the \actc\ data we updated pole \cite{sor} -- or
T-Model \cite{tmhi} -- HI (i.e. Higgs inflation)  by extending our
proposal in \cref{actpole}. Namely, we adopted the super- and \Kaa
potentials $W$ and $K$ in \eqs{whi}{tkis} which lead to the
inflationary potential in \Eref{Vhio} and the kinetic mixing in
\Eref{Je}. Moreover we required that the Higgs-inflaton assumes
ultimately its v.e.v predicted by the gauge unification within
MSSM. This requirement influenced via \Eref{Nmin} the parameter
space of \fhi\ and via \eqs{msn}{mrhi} the post-inflationary
scenario. Confining $(p,N)$ in the ranges shown in \Fref{fig2} we
achieved a nice covering of the present data -- see \Fref{fig1}.
Fixing $\ns$ to its central value in \Eref{data} we obtained the
clear predictions of \Eref{res1} regarding our free parameters. As
a consequence, our present framework drastically reduces the
allowed $(p,N)$ domain found for a gauge-singlet inflaton in
\cref{actpole}. It is gratifying that the resulting $r$ values for
$(p,N)$'s of order unity could be observationally accessible in
the foreseeable future. We also specified a post-inflationary
completion, based on the extra $W$ and $K$ terms in \eqs{dW}{dK},
which offers a nice solution to the $\mu$ problem of MSSM and
allows for baryogenesis via nTL with $\mrha$ in the range
$(10-76)~\EeV$ provided that $\mmgr\leq5.5$.

Since our main aim here was the introduction of \fhi, we opted to
utilize its embedding in a GUT based on $\Gbl$. Obviously our
scheme can be realized, employing the same $W$ and $K$'s, within
other SUSY GUTs too based on more structured gauge groups provided
that $\Phi$ and $\bar \Phi$ consist a conjugate pair of Higgs
superfields -- cf. \cref{hi5,hi4}. Special treatment is required
if the Higgs superfields belong to the adjoint representation of a
GUT -- cf. \cref{lrcs}. Moreover, given that the term $\ld M^2S/4$
of $W$ in \Eref{whi} plays no role during \fhi, our scenario can
be implemented by replacing it with $\ld'S^3$ and identifying
$\Phi$ and $\bar \Phi$ with the electroweak Higgs doublets $H_u$
and $H_d$ of the next-to-MSSM \cite{linde1}. In the last case a
tighter connection of the high- to the low-energy phenomenology is
expected.\\

\paragraph*{\small\bfseries\scshape Acknowledgments} {\small I would like to thank
F. Koutroulis, A. Masiero and Q. Shafi for interesting discussions
during the Corfu Summer Institute, where parts of this work were
presented.}

\appendix

\renewenvironment{subequations}{%
\refstepcounter{equation}%
\setcounter{parentequation}{\value{equation}}%
  \setcounter{equation}{0}
  \def\theequation{A\theparentequation{\small\sffamily\alph{equation}}}%
  \ignorespaces
}{%
  \setcounter{equation}{\value{parentequation}}%
  \ignorespacesafterend
}
\renewcommand{\thesubsection}{{\small\sf\arabic{subsection}}}

\section{Reheating Process, Lepton-Asymmetry and Gravitino
Abundances} \label{Rhg}

We present here a numerical, in \Sref{num}, and analytic, in
\Sref{an}, description of the post-inflationary evolution of our
setting justifying, thereby, the expressions adopted in
\eqss{Trh}{yb}{ygr}. From them the two latter results are
presented for first time, to our knowledge, in the literature.

\subsection{\sc\small\sffamily  Formulation of the Post-Inflationary Dynamics}
\label{num}

We display here the equations which govern the evolution of the
various energy and number densities involved in our scenario of
nTL. The initial form of these equations is given in \Sref{num1}
whereas a form more convenient for numerical manipulations is
given in \Sref{num2}.

\subsubsection{Initial Form} \label{num1}

The energy densities $\rho$ of $\dphi$, the energy density
$\rho_{\rm R}$ of the produced radiation, and the number densities
$n_L$ of the lepton asymmetry and $n_{3/2}$ of the $\Gr$'s satisfy
the following Boltzmann equations -- cf. \cref{turner}:
\beqs\begin{eqnarray}  &&\hspace*{-0.5cm} \dot
\rho+3(\wrh+1)H\rho+(\wrh+1)\Gsn
\rho=0,\label{rf}\\
&& \hspace*{-0.5cm} \dot\rho_{\rm R}+4H\rho_{\rm R}-(\wrh+1)\Gsn\rho=0,\label{rR}\\
&& \hspace*{-0.5cm}\dot n_{L}+3Hn_{L}-2\ve_{L}(\wrh+1)\GNsn\rho/\msn=0,\label{nL}\\
&& \hspace*{-0.5cm}  \dot n_{3/2}+3Hn_{3/2}-C_{3/2} n_{\rm
eq}^2=0.\label{ng}
\end{eqnarray}\eeqs
Here the overdot denotes derivation w.r.t the cosmic time $t$. Due
to the polynomial character of $\Vhi$ in \Eref{Vhio} and the
non-minimal kinetic mixing in \Eref{Je}, the estimation of $\wrh$
requires some care -- cf. \cref{rhlin}. We determine it adapting
the general formula \cite{turner}, i.e.
\beq w_{\rm rh}=2\frac{\int_{M}^{\sgm} d\sg\, J\, (1-
\Vhi/\Vm)^{1/2}}{\int_{M}^{\sgm} d\sg\, J\, (1-
\Vhi/\Vm)^{-1/2}}-1,\label{wrh}\eeq
where $\Vm=\Vhi(\sgm)$. The amplitude of the oscillations during
reheating $\sgm$ is found by solving numerically the condition
$\sqrt{3}\Hhi(\sgm)=\msn$ if $\sqrt{3}\Hhi(\sgf)>\msn$ or it is
$\sgm=\sgf$ otherwise. For $N=(0.1-8)$ where $J\simeq1$, $\wrh$
can be well approximated by performing an expansion for low $M$
values with the following result
\beq w_{\rm rh}\simeq \frac{1}{3}- 0.51 \frac{M}{\sgm}+0.22
\frac{M^2}{\sgm^2}+\cdots \label{wrh1}\eeq
which implies a reduction of $\wrh$ w.r.t its value ($1/3$)
corresponding to a pure quartic potential \cite{turner}. We also
checked that $\Vm\gg \Vhi(\sg=0)$ and so the oscillations about
the central maximum of $\Vhi$ (for $\sg=0$) can not be ignored.

Moreover, $n_{\rm eq}$ in \Eref{ng} is the equilibrium number
density of each bosonic relativistic species
\beq n_{\rm eq}= {\zeta(3)T^3/\pi^2}\eeq
and $C_{3/2}$ is a collision term for $\Gr$ production which, in
the limit of massless MSSM gauginos, turns out to be \cite{kohri,
brand}
\beq C_{3/2} = \frac{3\pi}{16\zeta(3)\mP^2}\sum_{\al=1}^{3} c_\al
g_\al^2 \ln\left(\frac{k_\al}{g_\al}\right),\eeq
where $(c_\al)=(33/5,27,72)$, $g_\al$ are the gauge coupling
constants of the MSSM -- $(g_\al(\Trh))=(0.54,0.68,0.88)$ for
$\Trh\sim\EeV$ -- and $(k_\al)=(1.634,1.312,1.271)$. Clearly, in
the limit of massless MSSM gauginos, the resulting $n_{3/2}$ is
practically $m_{3/2}$-independent.

Finally, the Hubble expansion parameter $H$ during this period is
given by
\begin{equation} \label{Hini}
H=\left(\rho +\rho_{\rm R} \right)^{1/2}/{\sqrt{3}\mP}.
\end{equation}
whereas the temperature $T$ and the entropy density ${\sf\small
s}$ are found from the relations
\begin{equation} \rho_{\rm R}=\frac{\pi^2}{30}g_*
T^4~~\mbox{and}~~{\sf\small s}=\frac{2\pi^2}{45}g_* T^3.
\label{rs}\end{equation}
The system of Eqs.~(\ref{rf}) -- (\ref{ng}) is solved under the
following initial conditions:
\beq\rho(0)=\Vhi(\sgm)~~\mbox{and}~~ \rho_{\rm
R}(0)=n_{L}(0)=n_{3/2}(0)=0.\label{init} \eeq
The results are obtained for a temperature $T_{\rm f}\ll\Trh$
where $\Trh$ is defined from the condition \cite{reh,reh1}
\beq\rho(\Trh)=\rho_{\rm R}(\Trh).\label{rehcon}\eeq
The abundances in \eqs{yb}{ygr} are estimated by the solution of
\eqs{nL}{ng} via the expressions
\beq Y_L=(n_L/{\sf\small s})(T_{\rm f})
~~\mbox{and}~~Y_{3/2}=(n_{3/2}/{\sf\small s})(T_{\rm
f}).\label{rhres} \eeq
%

\subsubsection{Reformulation} \label{num2}

The numerical integration of \Eref{rf} -- (\ref{ng}) is
facilitated, if we define the following variables -- cf.
\cref{reh}
\beqs \bea\label{fdef} &f=\rho R^{3(1+\wrh)},~f_{\rm R}=\rhoR
R^4,\\ &f_L=n_L R^3~~\mbox{and}~~ f_{3/2}=n_{3/2}
R^3,\label{fdefa}\eea\eeqs
so as to absorb the dilution terms. Indeed, \Eref{rf} --
(\ref{ng}) can be reexpressed in terms of the variables above as
follows
\beqs \bea \hspace*{-0.5cm}  H f' &=&-(\wrh+1)\Gsn f \label{ff},\\
\hspace*{-0.5cm}  H f'_{\rm R} &=&(\wrh+1)\Gsn f
R^{1-3\wrh},\label{fr} \\ \hspace*{-0.5cm}  H
f_{L}'&=&2\ve_L(\wrh+1)\GNsn f R^{3\wrh}/\msn,\label{fl} \\
\hspace*{-0.5cm}  H f_{3/2}' &=&C_{3/2} n_{\rm eq}^2 R^3,
\label{fg} \eea\eeqs
where prime denotes derivation w.r.t the logarithmic time
\beq \vtau=\ln\left(R/R_{\rm
i}\right)~\Rightarrow~R^\prime=R~~\mbox{and}~~R=R_{\rm i
}e^{\vtau}\label{vtau} \eeq
with $R_{\rm i}$ corresponding to the initialization of the
$\dphi$ oscillations. It can be conveniently selected so that the
resolution of the system from $\vtau_{\rm i}=0$  to $\vtau_{\rm
f}>\vtrh$ is numerically stable -- here $\vtrh$ corresponds to
$\Trh$.

\begin{figure}[!t]
\centering\includegraphics[width=60mm,angle=-90]{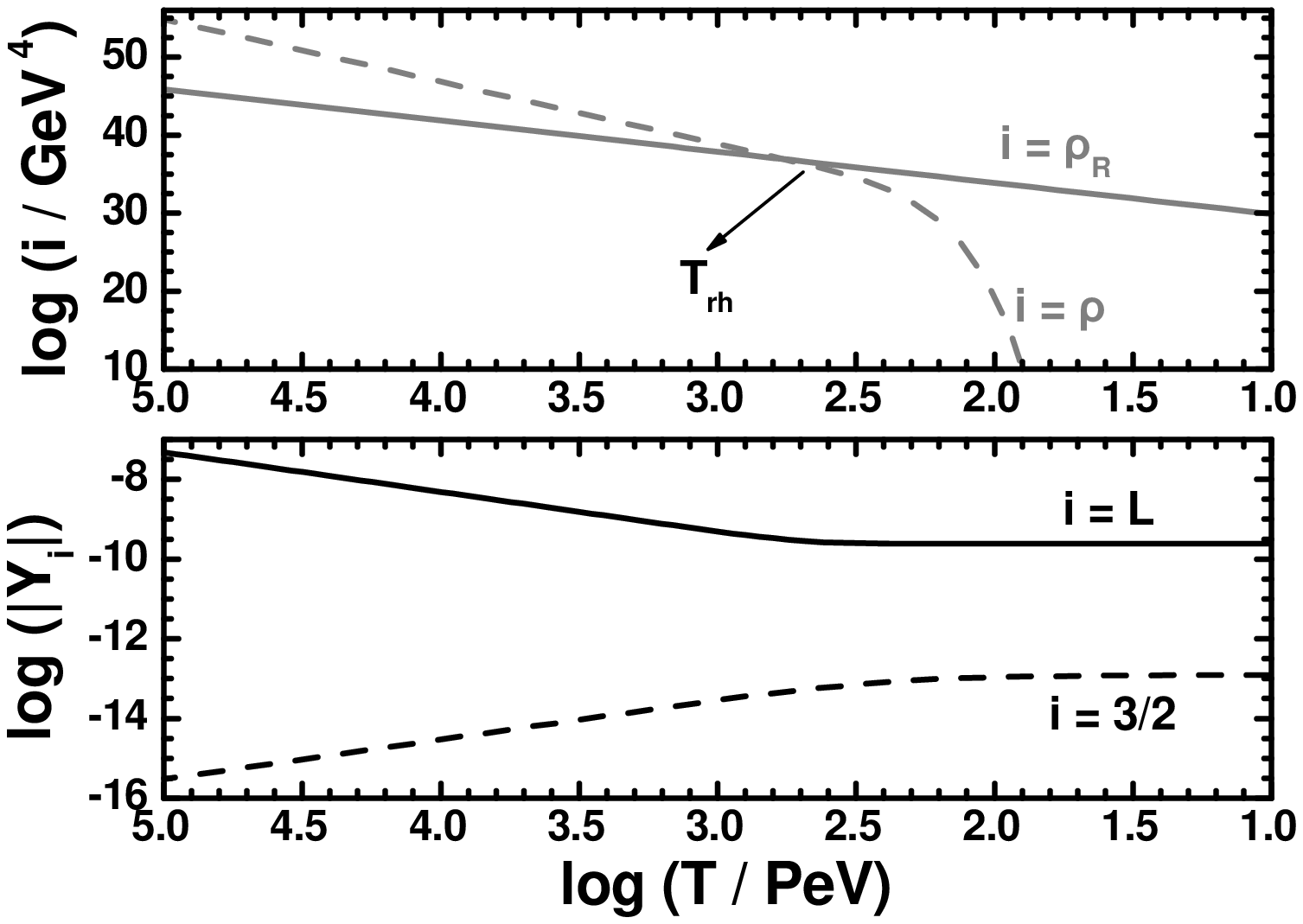}
\begin{ruledtabular}
\begin{tabular}{c|c|c|c|c|c}
Sec.~&$\vtrh~$&$\Trh~(\EeV)$&$H(\Trh)/\Gsn$&$\Yb~(10^{-11})$&$\Yg~(10^{-13})$
\\ \colrule
\ref{num}~&$14.3~$&$0.5$&$0.89$&$8.55$&$1.2$\\
\ref{an}~&$14.5~$&$0.57$&$0.87$&$9.6$&$0.51$\\ 
\end{tabular}
\end{ruledtabular}
\hfill\caption{\sl\small  The evolution of the quantities
$\log\rho$ (gray dashed line), $\log\rho_{\rm R}$ (gray line),
$\log |Y_L|$ (black solid line), and $\log \Yg$ (black dashed
line) as functions of $\log T$ for the inputs in \Eref{inp}. The
values of the resulting parameters found by employing our
numerical or analytic formulae presented in Sec.~\ref{num} and
\ref{an} respectively are listed in the Table.}\label{Trg}
\end{figure}

In Fig.~\ref{Trg}, we depict the evolution of the quantities
$\log\rho$ (dashed gray line), $\log\rho_{\rm R}$ (solid gray
line), $\log |Y_L|$ (black solid line) and $\log\Yg$ (black dashed
line) as functions of $\log T/\PeV$ for a representative set of
values of the parameters. In particular, we take
\beq N=0.3,~p=2,~\mmgr=0.1,~\mrha=33.5~\EeV. \label{inp}\eeq
We use also $\nst=1/3$ as in \Fref{frhn}. We obtain as outputs
\beqs\beq \Ns=55.9,~\ld=4.8\cdot10^{-5},~\msn=1.62~\ZeV,~\wrh=0.28
\label{outp1} \eeq
with the rightmost condition in \Eref{prob} being fulfilled for
$\sgx=0.9315\mP$. The resulting inflationary observables are
\beq \ns=0.9743,~\as=-4.7\cdot10^{-4}~~\mbox{and}~~r=0.014,
\label{outp2}\eeq\eeqs
whereas the outputs of the post-inflationary stage of our scenario
are listed in the Table of \Fref{Trg}. We there arrange values for
various quantities derived by our numerical approach described in
this section and by our analytical approximate expressions
presented in the next one. From the upper plot of \Fref{Trg} we
observe that \fhi\ is followed by a $\rho$-dominated era, due to
the oscillating and decaying $\dphi$, which lasts at the
intersection of the solid and dashed gray lines where $\Trh$ is
defined. After it, the universe enters a conventional radiation
dominated era. For $T=\Trh$ the black and dashed solid lines
approach their plateau (present) values as shown in the lower
panel of Fig.~\ref{Trg}.

\subsection{\sc\small\sffamily  Approximate Results}\label{an}

We focus on the regime with $T\geq\Trh$ where $\rho\geq\rho_{\rm
R}$ and therefore $H$ in \Eref{Hini} assumes the form
\beq H\simeq \sqrt{\rho}/\sqrt{3}\mP.\label{Hf} \eeq
Inserting it into \Eref{ff} we find
\beq \label{rfa} \rho\simeq\rhofi
e^{-3(\wrh+1)\vtau}~~\mbox{with}~~\rhofi=\rho(0)\eeq
given in \Eref{init}. Upon substitution in \Eref{fr} we obtain
\beq \label{rRa} \rho_{\rm R}(\vtau)=\rhoRi
\left(e^{-3(1+\wrh)\vtau/2}-e^{-4\vtau}\right),\eeq
where we define $\rhoRi$ as follows
\beq \label{rRi}
\rhoRi=\frac{2\sqrt{3}(\wrh+1)}{5-3\wrh}\mP\Gsn\rhofi^{1/2}.\eeq
The function $\rho_{{\rm R}}(\vtau)$ in \Eref{rRa} reaches a
maximum at
\beq \vtau_{{\rm
max}}\simeq\frac{2}{5-3\wrh}\ln\frac8{3(\wrh+1)},\eeq
which turns out to be very close to $\vtau=\vtau_{\rm i}$, e.g.,
$\vtau_{{\rm max}}-\vtau_{\rm i}=0.39$ or $0.35$ for $\wrh=0$ or
$\wrh=1/3$ respectively. Therefore, for $\vtau>\vtau_{{\rm max}}$
and $\wrh<5/3$, $\rho_{{\rm R}}$ in \Eref{rRa} is dominated by the
first term in the parenthesis. Upon substitution into the
rightmost relation in \Eref{rs} we find that $T$ decreases as
follows
\beq T=\Ti e^{-3(\wrh+1)\vtau/8}~~\mbox{with}~~\Ti=\lf
\frac{30\rhoRi}{g_*\pi^2}\rg^{1/4}. \label{Ttau} \eeq
For simplicity we take a constant value for $g_*$ throughout.

\subsubsection{Reheat Temperature}

The completion of the reheating takes place when \Eref{rehcon} is
satisfied. Taking into account \eqs{rfa}{rRa} we find that $\vtrh$
takes the form
\beq \vtrh\simeq
\frac{2}{3(\wrh+1)}\ln\frac{\rhofi}{\rhoRi}.\label{trh}\eeq
If we insert this result into \Eref{Ttau} and make use of
\Eref{rRi} we arrive at the result of \Eref{Trh} which does not
depend on $\rhofi$. Our approach is consistent with an alternative
definition of $\Trh$ according to which it is defined from the
condition $H(\Trh)\simeq\Gsn$. Indeed, we find
\beq H(\vtrh)/\Gsn=2\sqrt{2}(\wrh+1)/(5-3\wrh) \eeq
with numerical values given in the Table of \Fref{Trg}.


\subsubsection{Lepton-Asymmetry Abundance}

Integrating \Eref{fl} with $H$ given by \Eref{Hf} we find
\beq 
n_L=\frac{4(\wrh+1)\ve_L\mP}{\sqrt{3}(1-\wrh)\msn}\rhofi^{1/2}\Gsn
\lf e^{-3(\wrh+1)\vtau/2}-e^{-3\vtau}\rg,\label{nLa}\eeq
where the first exponential in the parenthesis dominates over the
latter for $\wrh<1$. Confining ourselves to this regime and
applying the expression above for $\vtau=\vtrh$ given by
\Eref{trh} we obtain
\beq
{n_L}(\vtrh)=\frac{8(\wrh+1)^2\ve_L\mP^2}{(1-\wrh)(5-3\wrh)\msn}\frac{\GNsn}{\Gsn}{\Gsn^2},
\label{nLb}\eeq
where $\rhofi$ is again eliminated if we take into account
\Eref{rRi}. Plugging into the expression above $\Gsn^2$ from
\Eref{Trh} and employing the rightmost expression in \Eref{rs} we
end up with the result of \Eref{yb} after including the numerical
factor $(-0.35)$ which accounts for the sphaleron effects.

\subsubsection{Gravitino Abundance}

Solving \Eref{fg} with $H$ and $T$ given by \eqs{Hf}{Ttau}
respectively we find
\beq n_{3/2}=\frac{4\zeta(3)^2\mP
C_{3/2}\Ti^6}{\sqrt{3}\pi^4\rhofi^{1/2}(3-\wrh)}\lf
e^{-3(\wrh+1)\vtau/4}-e^{-3\vtau}\rg.\label{n32a}\eeq
The first exponential in the parenthesis above is dominant for
$\wrh<3$. Working in that range of $\wrh$ values and replacing
$\vtau=\vtrh$ from \Eref{trh} we arrive at
%
%
%
%
\beq
n_{3/2}\lf\vtrh\rg=\lf\frac{10}{g_*}\rg^{1/2}\frac{4\zeta(3)^2\mP
C_{3/2}}{\pi^5(3-\wrh)}\Trh^4,\label{n32d}\eeq
where $\rhofi$ is again eliminated, as can be shown by replacing
$\Trh^2$ and $\rhoRi$ from \eqs{Trh}{rRi} respectively. Employing
the rightmost expression in \Eref{rs} we end up with
\beq \Yg=\frac{n_{3/2}}{\sf\small
s}\lf\vtrh\rg=\frac{90}{g_*}\lf\frac{10}{g_*}\rg^{1/2}\frac{\zeta(3)^2C_{3/2}}{\pi^7(3-\wrh)}\mP\Trh.
\label{n32c}\eeq
Comparing the result above with the numerical one -- see the Table
of \Fref{Trg} --, we remark that there is a slight discrepancy
which turns out to be systematic. Its origin stems from the fact
that there is a residual production of $\Gr$'s for $\vtau>\vtrh$.
To account for this effect we multiply the result of \Eref{n32c}
by a factor of $2$ and include it in the numerical factor of
\Eref{ygr} achieving, thereby, a better agreement with numerics.


\paragraph*{} In conclusion, we motivated the adopted formulas in
\eqss{Trh}{yb}{ygr}, which exhibit an explicit dependence on
$\wrh$ and allow us to analyze consistently the post-inflationary
dynamics. The numerical impact of this improvement on the results
is a correction of order unity and therefore it is not so crucial
for the viability of our set-up.


\def\ijmp#1#2#3{{\sl Int. Jour. Mod. Phys.}
{\bf #1},~#3~(#2)}
\def\plb#1#2#3{{\sl Phys. Lett. B }{\bf #1}, #3 (#2)}
\def\prl#1#2#3{{\sl Phys. Rev. Lett.}
{\bf #1},~#3~(#2)}
\def\rmp#1#2#3{{Rev. Mod. Phys.}
{\bf #1},~#3~(#2)}
\def\prep#1#2#3{{\sl Phys. Rep. }{\bf #1}, #3 (#2)}
\def\prd#1#2#3{{\sl Phys. Rev. D }{\bf #1}, #3 (#2)}
\def\npb#1#2#3{{\sl Nucl. Phys. }{\bf B#1}, #3 (#2)}
\def\npps#1#2#3{{Nucl. Phys. B (Proc. Sup.)}
{\bf #1},~#3~(#2)}
\def\mpl#1#2#3{{Mod. Phys. Lett.}
{\bf #1},~#3~(#2)}
\def\jetp#1#2#3{{JETP Lett. }{\bf #1}, #3 (#2)}
\def\app#1#2#3{{Acta Phys. Polon.}
{\bf #1},~#3~(#2)}
\def\ptp#1#2#3{{Prog. Theor. Phys.}
{\bf #1},~#3~(#2)}
\def\n#1#2#3{{Nature }{\bf #1},~#3~(#2)}
\def\apj#1#2#3{{Astrophys. J.}
{\bf #1},~#3~(#2)}
\def\mnras#1#2#3{{MNRAS }{\bf #1},~#3~(#2)}
\def\grg#1#2#3{{Gen. Rel. Grav.}
{\bf #1},~#3~(#2)}
\def\s#1#2#3{{Science }{\bf #1},~#3~(#2)}
\def\ibid#1#2#3{{\it ibid. }{\bf #1},~#3~(#2)}
\def\cpc#1#2#3{{Comput. Phys. Commun.}
{\bf #1},~#3~(#2)}
\def\astp#1#2#3{{Astropart. Phys.}
{\bf #1},~#3~(#2)}
\def\epjc#1#2#3{{Eur. Phys. J. C}
{\bf #1},~#3~(#2)}
\def\jhep#1#2#3{{\sl J. High Energy Phys.}
{\bf #1}, #3 (#2)}
\newcommand\jcap[3]{{\sl J.\ Cosmol.\ Astropart.\ Phys.\ }{\bf #1}, #3 (#2)}
\newcommand\jcapn[4]{{\sl J.\ Cosmol.\ Astropart.\ Phys.\ }{\bf #1}, #3, no.~#4 (#2)}
\newcommand\njp[3]{{\sl New.\ J.\ Phys.\ }{\bf #1}, #3 (#2)}

\end{document}